\documentclass[fleqn,10pt]{SelfArx} 

\usepackage[english]{babel} 

\usepackage{lipsum} 
\usepackage{float}
\usepackage{mathrsfs}

\definecolor{color1}{RGB}{0,0,90} 
\definecolor{color2}{RGB}{130,130,130} 

\usepackage{hyperref} 
\hypersetup{hidelinks,colorlinks,breaklinks=true,urlcolor=black,citecolor=color1,linkcolor=color1,bookmarksopen=false,pdftitle={Title},pdfauthor={Author}}

\JournalInfo{Preprint} 
\Archive{arXiv} 

\PaperTitle{Approaching the adiabatic timescale with machine-learning} 

\Authors{Bryce M. Henson\textsuperscript{a,1},
Dong K. Shin\textsuperscript{a},
Kieran F. Thomas\textsuperscript{b},
Jacob A.  Ross \textsuperscript{a},
Michael R. Hush \textsuperscript{c}
Sean S. Hodgman\textsuperscript{a},
Andrew G. Truscott\textsuperscript{a,1},
} 

\affiliation{\textsuperscript{a}\textit{Laser Physics Centre, Research School of Physics and Engineering, Australian National University, Canberra, ACT 2601, Australia}} 
\affiliation{\textsuperscript{b}\textit{School of Mathematics and Physics,University of Queensland, Brisbane, Queensland 4072, Australia}} 
\affiliation{\textsuperscript{c}\textit{School of Engineering and Information Technology, University of New South Wales at the Australian Defence Force Academy, Canberra, 2600, Australia}} 
\affiliation{\textsuperscript{1}\textbf{Corresponding authors}: Bryce.Henson@anu.edu.au, Andrew.Truscott@anu.edu.au} 

\newcommand{\keywordname}{Keywords} 

\Abstract{The control and manipulation of quantum systems without excitation is challenging, due to the complexities in fully modeling such systems accurately and the difficulties in controlling these inherently fragile systems experimentally.  For example, while protocols to decompress Bose-Einstein condensates (BEC) faster than the adiabatic timescale (without excitation or loss) have been well developed theoretically, experimental implementations of these protocols have yet to reach speeds faster than the adiabatic timescale.
In this work, we experimentally demonstrate an alternative approach based on a machine learning algorithm which makes progress towards this goal.
The algorithm is given control of the coupled decompression and transport of a metastable helium condensate, with its performance determined after each experimental iteration by measuring the excitations of the resultant BEC. After each iteration the algorithm adjusts its internal model of the system to create an improved control output for the next iteration. 
Given sufficient control over the decompression, the algorithm converges to a novel solution that sets the current speed record in relation to the adiabatic timescale, beating out other experimental realizations based on theoretical approaches. 
This method presents a feasible approach for implementing fast state preparations or transformations in other quantum systems, without requiring a solution to a theoretical model of the system.
Implications for fundamental physics and cooling are discussed.
}
\newcommand{\LaySumName}{Significance} 
\LaySum{Engineering the fast evolution of a quantum system between states is a key problem to be solved in the development of quantum technologies, such as quantum computing. 
We experimentally demonstrate a general approach using a Machine Learning algorithm that develops a model of the system, based on previous performance, to create further educated guesses on how to improve.
Applied to a system similar to moving a cup of liquid between two locations (while blindfolded) the algorithm reaches a speed faster than previous approaches, dealing well with the complex dynamics and experimental imperfections present with its empirical approach. The resulting fast dynamics open the door to understanding how quantum mechanical systems reach equilibrium while the method provides a new tool to taming complex quantum systems.
}

\begin{document}

\flushbottom 

\maketitle 


\thispagestyle{empty} 


\section*{Introduction} 

The precise control and preparation of quantum states is a cornerstone in the quest for quantum computers and quantum communication systems. 
A particularly relevant example is state preparation: controlling the transformation of a system from the ground state of one potential to that of another through its evolution under a limited control Hamiltonian.
A common approach here is the adiabatic one where the engineered evolution in the Hilbert space is so slow that there is a negligible probability of exciting other (instantaneous) eigenstates throughout the transformation.
In the absence of decoherence and loss these protocols work well. However, real systems suffer from these corrupting effects which compound with longer evolution, making such approaches sub-optimal.
Additionally these protocols are often impractically slow to implement for proposed quantum devices \cite{PhysRevB.96.075158}.
The solution to this problem is to engineer a faster evolution that results in a final state $\lvert\psi_\textrm{f} \rangle$ in some finite control time $\tau$ that approximates or is equal to the state produced by the infinitely slow ($\tau\rightarrow\infty$)  adiabatic process $\lvert\psi_\infty \rangle$, such that $\lvert\langle \psi_\infty \vert\psi_\textrm{F} \rangle\rvert^2 \approx1 $. 
These approaches have been investigated theoretically under the terms \textit{shortcuts to adiabaticity} \cite{TORRONTEGUI2013117,Chen2010review,Schaff2011,Chen2010}, \textit{counterdiabatic driving} \cite{PhysRevA.97.040302} and \textit{optimal control} \cite{PhysrevLett.106.190501,PhysrevLett.103.240501,PhysRevLett.89.188301}. 
While promising unprecedented performance, these theoretical protocols cannot be easily implemented in practice due to the complexities involved in real systems \cite{SchaffSongCapuzziEtAl2011,Schaff2011,PhysRevA.92.043416} which have two effects. 
First, any model  used  to solve an optimal control problem invokes approximations that may or may not capture all relevant dynamics in a given system. 
Second, idealized control inputs are difficult or impossible to deliver to a system given the variability and noise inherent in the transfer functions of optical, electronic, and mechanical transducers.

An approach to circumvent this is to find control parameters that give the optimal state preparation by empirical optimization, without necessarily knowing the details of how this control acts on the system  \cite{0953-4075-40-18-R01,5676677}.  
This method, originally proposed to excite particular molecular states \cite{PhysRevLett.68.1500} is a relatively simple concept that has a been studied for a range of systems such as quantum gates \cite{PhysRevLett.112.240503,PhysRevLett.112.240504}, isomerization  \cite{doi:10.1063/1.3103486}, the evaporative cooling of neutral atoms \cite{Wigley2016}, and the optimization of magneto-optical traps \cite{2018arXiv180500654T}.
Most previous works on automatic optimization of quantum science experiments have used relatively unsophisticated methods, such as evolutionary algorithms \cite{Judson1992,Warren1581,Baumert1997,Rohringer2008,Geisel2013}, while a novel approach of `gamification' allowed citizen scientists utilizing human intuition to produce solutions to complex quantum problems \cite{Sorensen2016} without understanding the underlying theory.
These approaches can take an excessive number of experimental runs to converge to an optimal solution, as they do not use all available information to determine the next best point to test. 

\begin{figure}[t]
\centering
\includegraphics[width=\linewidth]{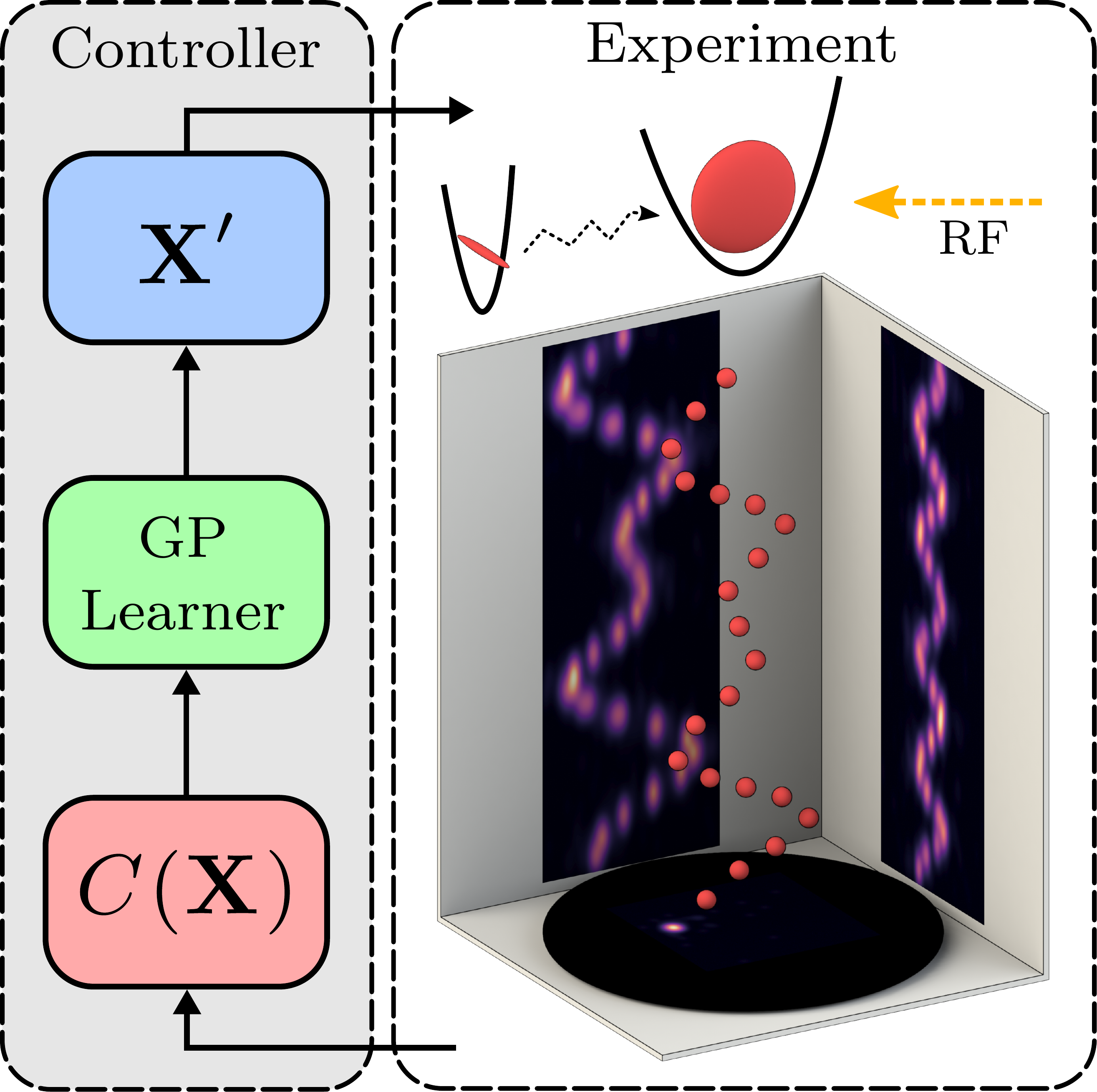}
\caption{
Experimental schematic of the transport and decompression of the condensate from the initial tight configuration (upper left) to the final relaxed trap (upper right). 
Excitations present in the end are probed by a series of RF pulses which uniformly transfer small fractions of atoms into the un-trapped states (see \textit{Materials and Methods}).
These atoms then fall freely under gravity and are detected individually with 3D spatio-temporal resolution (bottom).
The cost function $C(\mathbf{X})$ is calculated from the measured temporal evolution of the atoms for the current set of experimental parameters $\mathbf{X}$, and returned to the GP learner.  
The GP learner then generates a new iteration of the control parameters $\mathbf{X'}$ based on the measured cost functions of previous experimental parameters, which is then used in the next experimental sequence. 
Note that the relative sizes of the trapped BECs before and after the transformation are drawn to scale, while the displacement has been reduced ten-fold.
}
\label{fig:fig1}
\end{figure}

In this paper, we use a machine learning (ML) algorithm to optimize the complex problem of controlling the transformation of a quantum state, wherein we relax the harmonic confinement, decreasing the trapping frequencies (decompression), and simultaneously spatially translate the mean position (transport) of a Bose-Einstein Condensate as shown in Fig. \ref{fig:fig1}. 
This control problem is particularly difficult since the BEC's superfluidity causes excitations to persist far longer than the experimental sequence duration.
In addition, the relationship between the control parameters used (inputs to trap current controller) and their influence on the system is complex and non-linear, making this a challenging problem 
to solve analytically. 

Such transport-decompression schemes are commonly used in trapped ion \cite{PhysRevLett.109.080502} and ultra-cold atom experimental systems \cite{Khakimov2016}.
They are particularly important for BEC experiments, since the production of BEC with the standard evaporative cooling technique is most efficient in tight traps, which are then modified for different experiments (e.g. \cite{Khakimov2016}).
We let a ML algorithm control the transformation between well defined start and endpoints to prepare a BEC in the required relaxed trap in the shortest time achievable. 
In a separate experiment, we excite center of mass (COM) oscillations in the BEC and entrust the ML algorithm with similar experimental control parameters to remove the excitation.

There has been a series of demonstrations using ML techniques for the optimization of a range of quantum \cite{PALITTAPONGARNPIM2017116,PhysrevA.95.012335,Palittapongarnpim1062341,1705.00565,1704.06514,Li2017,PhysrevX.7.021024,Mavadia2017} and classical systems \cite{doi:10.1080/03052150500211911,doi:10.1080/03052150600882538,1299853,1331084}. 
The ML algorithm used in this work is based on the Gaussian process (GP) \cite{rasmussen2006gaussian}.  This builds a model of the cost function, reflecting the goodness of the experimental outcome with respect to the input control parameters, and learns by iteratively improving the model with the automated acquisition of new data points.
For the implementation we used an open source software \textit{M-LOOP} \cite{Mloop}, first used to automatically optimize the evaporative cooling stage of an ultra-cold atom experiment \cite{Wigley2016}.

\section*{Optimal Control and Adiabatic Timescales}
For a classical particle in a harmonic potential, driving of particle motion is suppressed for driving frequencies away from the resonant frequency. 
Therefore, the adiabatic condition is simply that the power spectral density of the time-varying potential vanishes around the (instantaneous) resonant frequency of the oscillator \cite{PhysRevLett.109.080502,CouvertKawalecReinaudiEtAl2008}. 

However in the quantum mechanical case, the problem is more subtle. 
The first adiabatic approximation, as stated by Born and Fock in 1928 \cite{BornFock} and proven by Kato in 1950 \cite{Kato}, is that
\begin{align}
\label{H_gap}
\max_{t\in [0,t_f]} \frac{|\langle{\phi_m}\lvert\partial_t{H}\rvert\phi_n\rangle|}{|E_n-E_m|^2} \rightarrow  0,
\end{align}
where $\vert\phi_i(t)\rangle$ are instantaneous eigenstates of energy $E_i$ of the time-varying Hamiltonian $\hat{H}$. The general adiabatic theorem requires careful treatment, and is still subject to examination\cite{Albash}.
The slow variation of $\hat{H}$ is more restrictive than the classical case, as the smoothness of $\hat{H}$ implies a narrow spectrum, hence constraining the transformation to a timescale $\tau < 2\pi/\omega_{f}$, where $\omega_{f}$ is the lowest characteristic frequency of the final trap configuration. 
The condition Eq.~\ref{H_gap} places stronger restrictions on $\dot{H}$ as the gap $E_n-E_m$ decreases, and the spectrum of a superfluid BEC is gapless which further complicates the task as the condensate is particularly susceptible to heating. 

Another challenge is the increased sensitivity of the transformation performance as the transformation time decreases. 
The probability of transition out of the initial eigenspace for a gapped system during a process of length $\tau$ scales as $\mathcal{O}(1/g^2 \tau^2)$, where $g$ is the minimum energy gap between the initial eigenspace and all other eigenspaces \cite{Jansen}. 
The transition probability is reduced for every order that the Hamiltonian's evolution is differentiable,
which ultimately sets the limit achieved in this work: with a limited number of control parameters, $\hat{H}(t)$ can never be arbitrarily smooth and hence performance is limited.

For the problems of either linear translation or variation in characteristic frequency, previous approaches employed dynamical invariants to solve for a fast trajectory that yields a final state identical to the final state of an adiabatic transformation \cite{Chen2010,PhysRevA.89.063414,SchaffSongCapuzziEtAl2011,Schaff2010}. 
Our simultaneous deformation and translation, along with coupling between axial modes of our trap make this problem particularly difficult.  While this is theoretically not a limit, the model-free approach that we demonstrate here could be generalized to more complicated systems where the model is intractable or even unknown.

Equation~\ref{H_gap} can be translated into specific contexts to provide so-called `adiabatic timescales', which are not necessarily timescales realizable in an experiment, but the ultimate limitation on the rate of a transformation before transient excitations are absolutely unavoidable. 
In the case of decompression of a one-dimensional quantum harmonic oscillator, the adiabatic condition is $\vert\sqrt{2}\dot{\omega}/8\omega^2 \vert \ll 1$, where $\omega(t)$ is the instantaneous oscillator frequency \cite{Chen2010}. 
The adiabatic timescale $T_{(ad)}$ can be found by solving for the equality in the adiabatic condition, which gives
\begin{equation}
T_{(ad)}=\frac{\omega_i-\omega_f}{4 \sqrt{2}\omega_i\omega_f},
\label{eq:ad_timescale_decompression}
\end{equation}
where $\omega_i$ and $\omega_f$ denote the initial and final oscillator frequencies.
To our knowledge the fastest decompression of a harmonic trap to date is slower than the adiabatic timescale by a factor of $\sim18$ \cite{PhysRevA.97.013628,SchaffSongCapuzziEtAl2011}, although factors of $\sim10^3$ are often used \cite{Leanhardt1513}. 

For the experimental conditions used in the decompression described in this work $\omega_i = 2\pi\cdot 595$~Hz, $\omega_f = 2\pi\cdot 5.8$~Hz along the radial axis, for which the adiabatic timescale is $T_{ad}=4.8$~ms.

\begin{figure}[t]
\centering
\includegraphics[width=\linewidth]{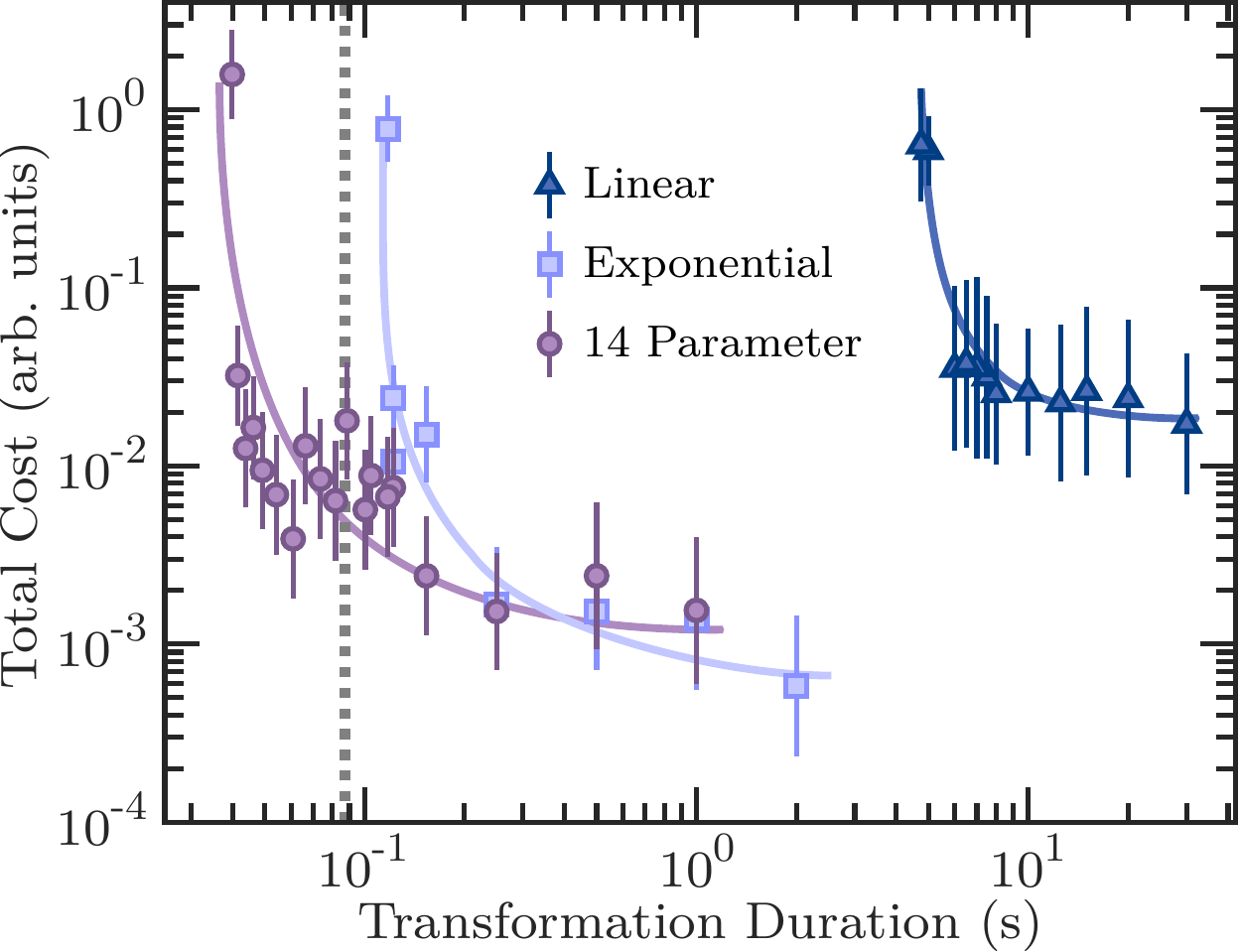}
\caption{
Minimum cost functions, which are a measure of the superfluid excitations, after optimization by the ML algorithm for different total transformation times and forms.  
Triangles show simple linear current ramps between the initial and final traps, which are seen to always induce some excitations that become severe heating below $\sim$ 6s.
An exponential current ramp (squares) with 2 control parameters performs better, with only minimal excitations for long control durations, although as the duration is decreased below 200ms this also fails. 
Circles show the results for 16 linear current segments with 14 control parameters, which performs much better as the algorithm now has sufficient freedom to control the transport and decompression without adding significant energy to the system.  For extremely short control durations $\lesssim$ 50ms this approach also fails, as the ideal control is badly approximated.  
Lines are shown as guides to the eye.
The vertical dashed line shows the previous best effort at $\sim 18 \times$ the adiabatic decompression timescale.
For a breakdown of the component cost see SI Fig.~\ref{fig:fig_si3}.}
\label{fig:fig2}
\end{figure}

\begin{figure}[tb]
\centering
\includegraphics[width=\linewidth]{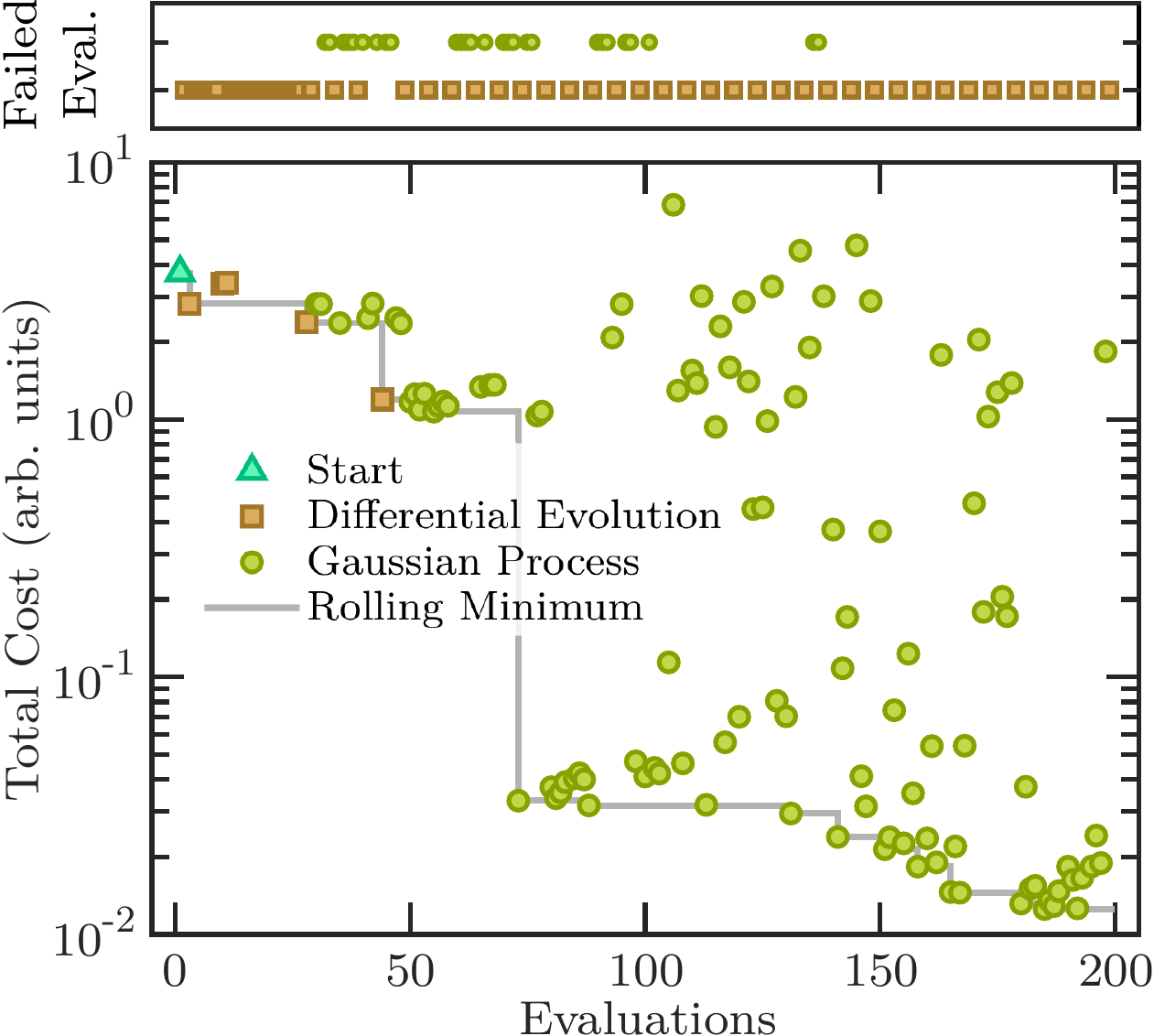}
\caption{
The measured cost functions for the optimization sequence of the 43ms control duration with 14 parameters, corresponding to 16 linear current segments.  
Each point indicates an experimental realization of a BEC and subsequent transformation.  
The upper row indicates experiments that were deemed to 'fail' as the cost function could not be evaluated due to insufficient numbers of detected atoms.  
The circles indicate realizations using parameters chosen by the Gaussian process in the ML algorithm, while the squares show those where the parameters were determined from the differential evolution method (see \textit{Materials and Methods} for details). 
Over time the best observed cost function (solid line) is seen to decrease, indicating that the algorithm is converging on a solution, although in a somewhat erratic manner due to the complexity of the cost landscape. The later failure of differential evaluation realizations (which are more 'exploratory' probes of parameter space) indicate a narrow optima is reached.}
\label{fig:fig3}
\end{figure}

\section*{Experiment}

The experiment starts with a BEC of $N\sim 8 \times 10^5$ He* atoms in a magnetic trap with no discernible thermal fraction (for more details see \textit{Materials and Methods}).
Initially, the trap is cigar shaped with $\left( \omega_x,\omega_r \right) \approx 2\pi \cdot \left(52,595 \right)\textrm{Hz}$. 
The trap is then decompressed by changing the current through  two pairs of coils \cite{Dall2007a} until it forms a much more weakly confined, pancake shaped harmonic trap with $\left( \omega_x,\omega_r\right) \sim 2\pi \cdot \left(5.8,15.0\right)\textrm{Hz}$.
Due to the coil geometry, the final trap has an inverted aspect ratio compared to the initial trap and the trap center also moves by $\approx9$~mm in the process. 
Complicating the transport processes is the small potential depth of the final trap corresponding to an escape velocity in the x direction of $\sim 200$~mm/s, limiting the maximum achievable acceleration imparted by the transformation \cite{PhysRevA.97.040302}.

To measure how well the final state agrees with a slow transformation (adiabaticity),  we characterize the excitations of the BEC after the transport and decompression, which include 3D sloshing and breathing modes, along with heating of the condensate.
This is possible as we are able to measure time-resolved 3D momentum distributions of the atoms over many trap oscillation cycles after the transformation (see \textit{Materials and Methods} for details).
The ML algorithm determines the performance of an experiment by evaluating a cost function which depends on the amplitude of the excitations.  A new set of parameters for the next experimental sequence is then generated based the measured cost functions for all previously tested experimental parameters.  
The experimental procedure is shown schematically in Fig.~\ref{fig:fig1}.
For each transformation duration (i.e. each data point in Fig. \ref{fig:fig2}) the algorithm adjusts the free ramp parameters over multiple experimental runs, using the optimal ramp parameters found from the previous (slower) transformation as initial conditions.
This convergence process for a single ramp is shown in Fig.~\ref{fig:fig3}.

\section*{Optimization}

A number of different forms of the transport and relaxation transformation are used for comparison. 
For a reference point, we start with a linear current ramp for both coil sets with no free parameters.
The cost functions for these linear ramps are shown as triangles in Fig.~\ref{fig:fig2}.  
Unsurprisingly, the linear ramps result in extreme excitations of the BEC for everything except the longest ramp durations, and even then the final state is far from equilibrium.  

As a first attempt at optimization, we use the algorithm to optimize an exponential ramp between the initial and final coil currents $I(t)=I_f + \exp(-t/\tau_c)\left( I_i - I_f \right)$ for each pair of coil currents, with the free parameters being the time constant $\tau_c$ of the exponential function.  
The results are shown for various ramp durations as squares in Fig.~\ref{fig:fig2} and show that the exponential ramps perform significantly better than the simple linear case.  
At long ramp durations, there is almost no excitation in the final state.  
However, for ramp durations shorter than $\sim 200$~ms, which is still $\sim 50$ times greater than the adiabatic limit, significant excitations are observed even in the optimized transformations. 
Further analysis of the cost function breakdown shows that trend comes predominantly from the heating term (SI Fig.~\ref{fig:fig_si3}).

We then give the algorithm greater freedom, allowing it to optimize the current ramp through the two pairs of coils, with each ramp consisting of eight linear segments between a fixed start and end point. 
For simplicity the duration of each individual segment is kept constant and equal to all other segments, with only the start and end points of each ramp used as free parameters to optimize.  
The optimized cost for these two eight segment ramps (14 free parameters total) are shown as circles in Fig.~\ref{fig:fig2}.  As can be seen, this ramp type performs better than the exponential case at short durations, and is able to reach approximately 9 times the adiabatic timescale before the excitations become severe. 
This is closer to the adiabatic timescale than has ever been previously reached, see Ref.~\cite{PhysRevA.97.013628}, by a factor of 2.

\begin{figure}[t]
\centering
\includegraphics[width=\linewidth]{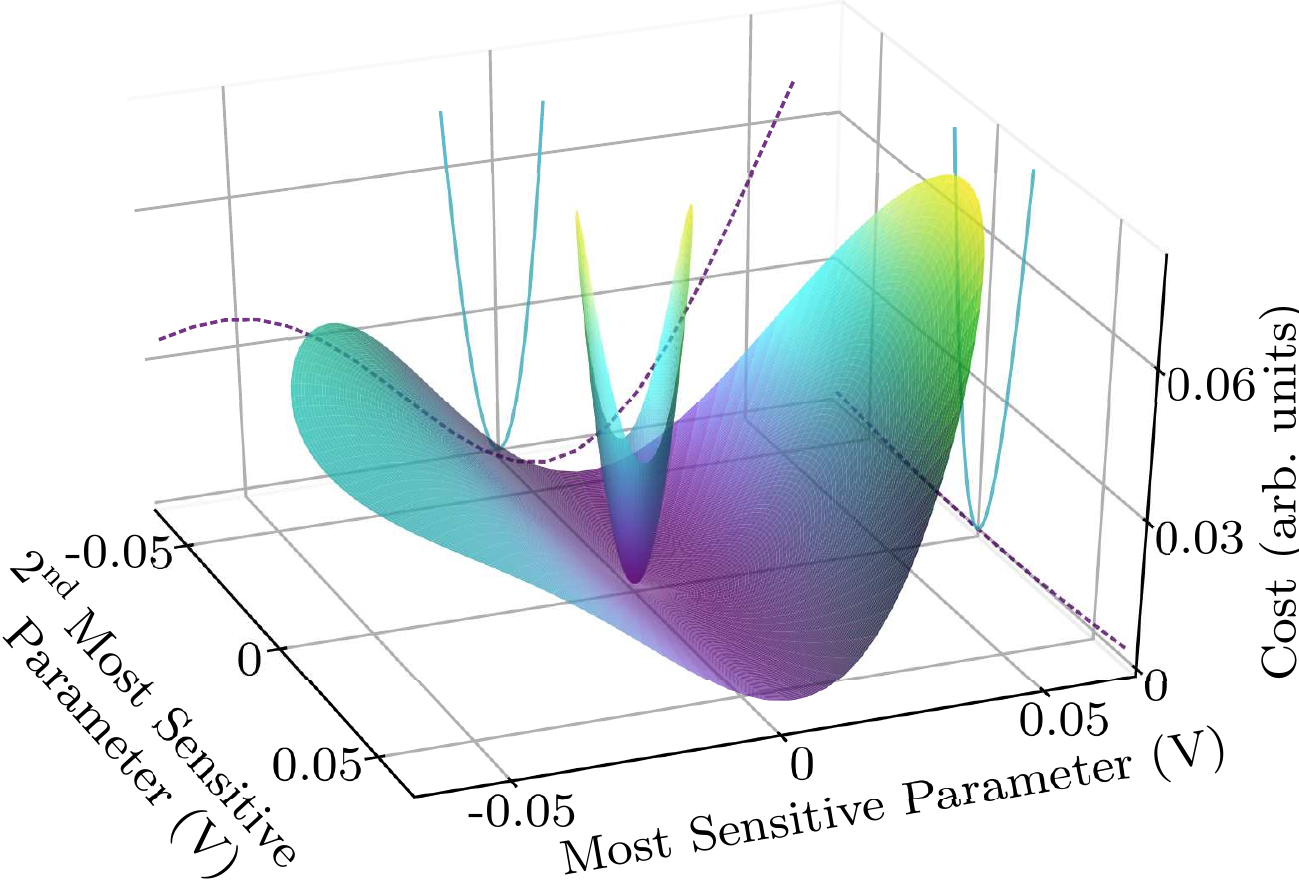}
\caption{A two dimensional region of the cost landscape around the optimized values of the most and second most sensitive parameters, as developed by the Gaussian process model for transformation durations of 43~ms (narrow function, blue solid line) and 1000~ms (broad function, dashed purple line) with the 14 parameter control linear segment transformation. The two most sensitive parameters are determined by the second derivative about the optima. 
The optima is subtracted for each control parameter.}
\label{fig:fig4} 
\end{figure}

From the work of \cite{PhysRevA.89.063414} it is known that schemes increase in sensitivity to both noise and systematic offset as the speed increases. 
To this end we perform an analysis of the cost landscape produced by the GP algorithm comparing the 2D landscape produced by the most sensitive parameters for both the 43 and 1000~ms control duration, see Fig. \ref{fig:fig4}. 
If we characterize the sensitivity by the curvature of the surface about the optima, the 43~ms transfer is $\sim 3500$ times more sensitive. 

\begin{figure}[t]
\centering
\includegraphics[width=\linewidth]{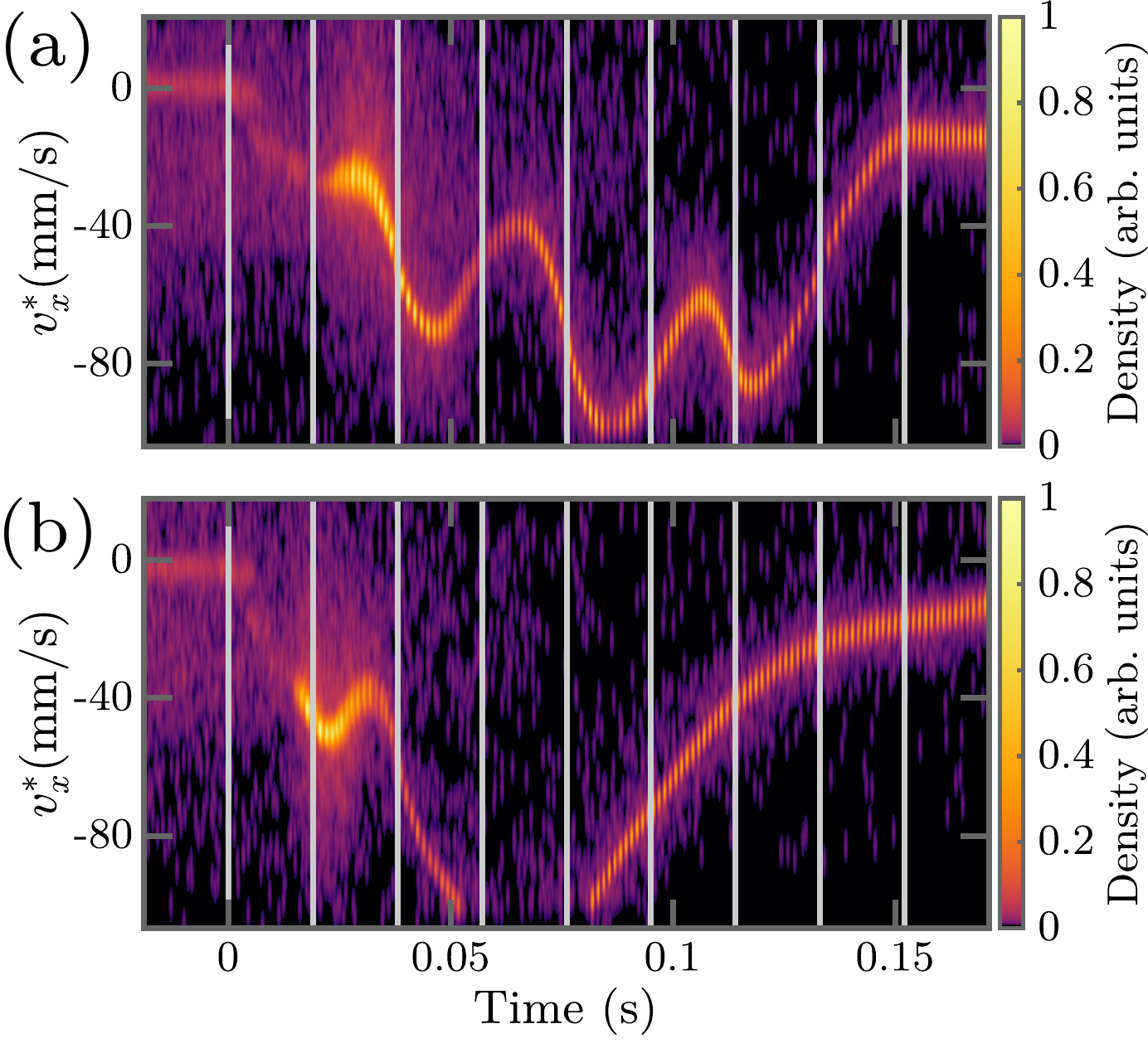}
\caption{
The measured far-field velocity ($v^{*}_x$) distributions over time along the weak trap axis during the ML optimized transport ramps produced for 153.6~ms total ramp duration. 
As the atoms are detected at a finite (852~mm) distance below the BEC,
the measurement is a combination of position and velocity $v^{*}_x=v_x+x/t_{\textrm{fall}}$.  
The displacement in the trap position following the transformation can be seen by the small difference in the start and end values.
During the optimized 14 linear segment ramp (a) the algorithm appears to induce an oscillation through a rapid transformation, which is then canceled out with the following current ramp.
This is repeated four times, resulting in a rapid transformation with minimal induced oscillations in the final state. 
(b) In contrast, the optimal exponential ramp for the same parameters also induces  oscillations, but due to a lack of adjustable parameters is unable to sufficiently cancel them out, resulting in substantial oscillations in the final state (not visible). 
The measured cost functions for these cases is $2.4\times10^{-3}$ and $1.5\times10^{-2}$ for (a) and (b), respectively. 
For the current waveforms applied to the trap for these transformations see SI Fig.~\ref{fig:fig_si1}. 
A linear transformation for this time was unable to produce atoms in the final state.}
\label{fig:fig5}
\end{figure}

To gain more insight into the dynamics of the BEC during the transformation and the strategies used by the ML algorithm to optimize the transformation, we use a modified outcoupling scheme to probe the in-trap momentum during the transformation (see  \textit{Materials and Methods} for details).  
The measured momentum evolution along the $\hat{x}$ direction (corresponding to the weak axis of the initial trap) for the optimized 153.6~ms duration 14 parameter ramp is shown in Fig.~\ref{fig:fig5}(a).
The algorithm appears to achieve a small final oscillation amplitude by repeatedly `throwing' and `catching' the atoms during the ramp. 
This appears to be similar to the bang-bang control schemes which have been shown theoretically to be the optimal control solution in problems where the maximum energy is bounded \cite{B816102J,PhysRevA.84.043415} and have been employed in experimental schemes \cite{PhysRevA.92.043416}.  
To compute such a scheme manually would be difficult due to the anharmonicity of the trap at large distances from the trap center, along with the losses that can occur due to the finite depth of the trap.  
This is therefore an example of a ML algorithm discovering an optimal transformation scheme that would be extremely challenging to design or optimize manually.  

In contrast, the velocity during the optimized exponential ramp for the same total duration (153.6~ms) is shown in Fig.~\ref{fig:fig5}(b).  
As can be seen, with fewer parameters to adjust the algorithm is unable to compensate for or prevent the oscillation that is excited during the ramp.  
As a result the final state displays large amplitude oscillations, which gives a higher cost function.

To improve the transformation further, with the ultimate aim of reaching the adiabatic limit, it would be beneficial to increase the number of parameters available to control, for example by increasing the number of parameters in the linear ramp or by making the length of each segment an adjustable variable. 
However, this comes at a substantial cost in the convergence and calculation times of the ML model, which has to re-calculate the model between experimental runs and feed back the new parameters before the next shot.  
This practically limited the maximum number of parameters possible to control in this work to 14, although improvements may be possible (see SI).

\begin{figure}[t]
\centering
\includegraphics[width=\linewidth]{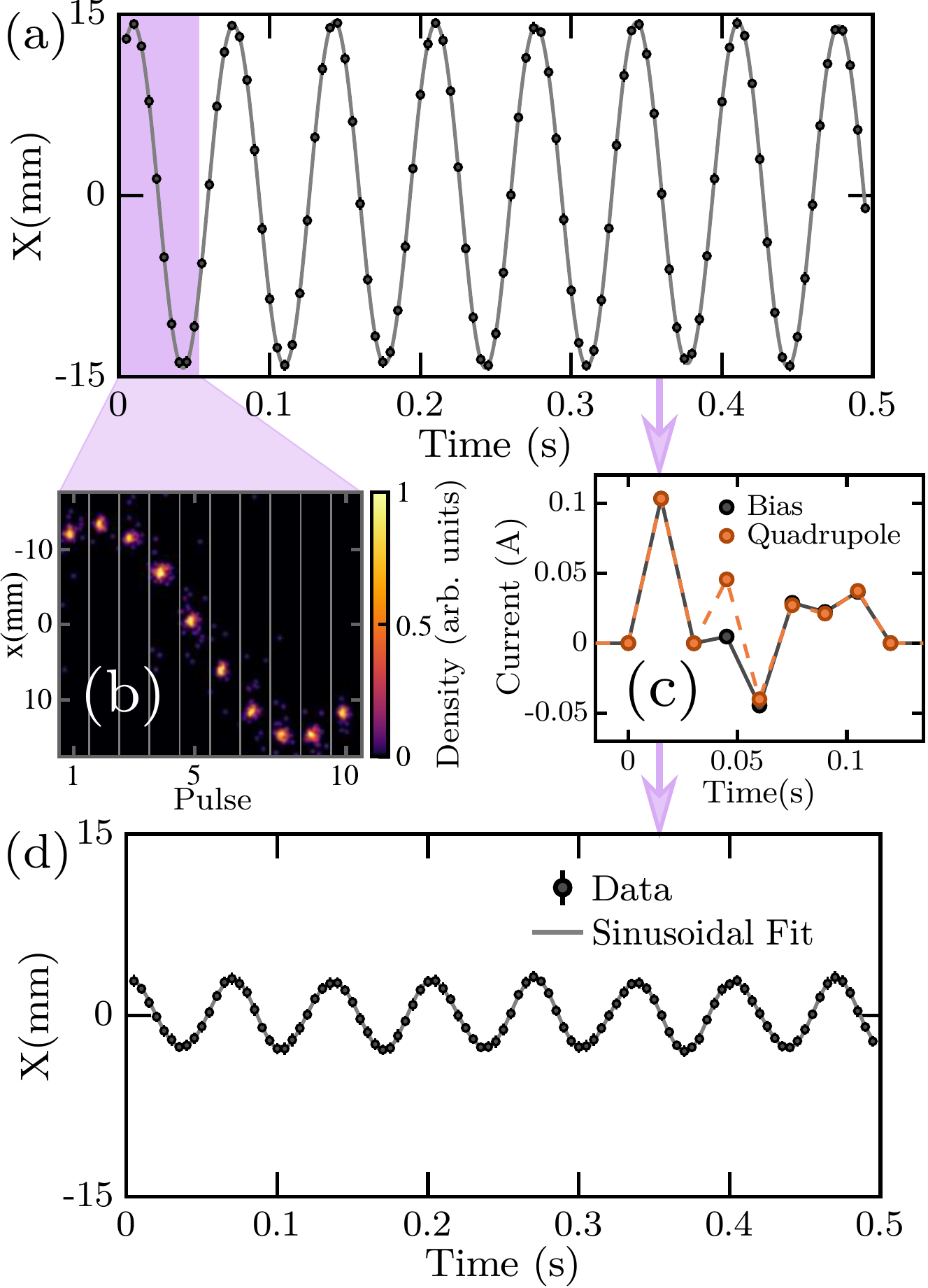}
\caption{
Fast damping of motion. 
Large oscillations (a) are induced in the motion of the condensate by a non-adiabatic ramp. 
 The individual atom laser pulses are shown in (b), with different time slices separated by vertical bars.  This unwanted motion is subsequently attempted to be removed by the algorithm through the applied control of 8 linear current ramps through the two pairs of trap coils (c), resulting in reduced oscillations (d).
Error bars are smaller than plot markers. 
Grey lines show sinusoidal fits to the data, with amplitudes of 14.20(1)~mm, 2.82(1)~mm for (a) and (b) respectively.
}
\label{fig:fig6}
\end{figure}

As a further test of our method we investigated the possibility of optimal damping of the system following an imperfect transport. 
For this experiment the final trap has stronger confinement with a  trap frequency of $(\omega_x,\omega_r) \approx 2\pi (15,25)\textrm{~Hz}$ and is `deeper', with a higher escape velocity of $\sim 500$~mm/s.
The system first undergoes a non-optimal linear transformation over 1200~ms and we then give the optimization algorithm control of the trap current for 120~ms, with the values of eight equal duration linear current segments per coil pair providing 14 free parameters. 
We then measure performance using the pulsed outcoupling scheme described above and the same cost function. 
Our results (Fig. \ref{fig:fig6}) show that the optimizer is able to reduce the cost function by a factor of $\sim$4 (from 0.01 to 0.0035), and the corresponding energy in the COM oscillation by a factor of $\sim$8 in a mere 1.8 trap cycles for the $x$ axis after a few hundred experimental realizations. 
This technique provides a demonstration that energy can be quickly removed from oscillations of a BEC, a first step towards active feedback cooling of a BEC \cite{PhysRevA.82.043632,1367-2630-15-11-113060,PhysRevA.87.013626,0953-4075-46-8-085301}. 
Further, it shows that the transport we perform is not as simple as the transport of a classical particle, which would allow for arbitrarily efficient damping.

\section*{Conclusion and Outlook}

In conclusion, we have demonstrated a significant improvement in the rapid transformation of a quantum system when the transformation parameters are optimized by a ML algorithm.  
The algorithm was able to speed up transformation substantially while preventing heating, atom loss, and COM oscillation, with the optimal results approaching the adiabatic limit.  
Additionally, when an oscillation of the condensate was deliberately excited, this oscillation could be substantially reduced by a similar ML algorithm.  

The technique we demonstrate here of using ML algorithms to improve transport in a quantum system can be generalized to a number of other problems.  
This novel method of rapid transformation could be used to generate a large, rapid change in the system Hamiltonian (a strong quench) without adding excess energy to the system for the study of non-equilibrium physics \cite{edelman2017chaotic}.
Applying the optimization technique to cooling could allow fast cooling below the limits generally reached in cold atom systems \cite{Leanhardt1513}.   
Applied to many-body lattice systems this could be particularly interesting, as high temperatures have thus far prevented the study of long sought after low temperature states such as \textit{d}-wave superfluidity \cite{Esslinger2010}.  
More generally, the techniques demonstrated here are likely to be useful for more broader applications to quantum technologies, such as the optimization of quantum gates.
\section*{Materials and Methods}

\subsection*{Magnetic Trap and BEC Transport}
The magnetic trap consists of two sets of anti-Helmholtz coils in a bi-planar quadrupole Ioffe configuration \cite{Dall2007a}. 
The first pair (\textit{quadrupole}) generates the dominant quadrupole potential for trapping neutral atoms, while the second pair (\textit{bias}) produces a non-zero bias field in the \textit{Ioffe} configuration.
We initially prepare the BEC in the \textit{tight} cigar-shaped trap with $\{\omega_x,\omega_r\}\approx2\pi\{52,595\}\textrm{Hz}$ (as in our previous work \cite{Dall2007a,Hodgman2017}) corresponding to $I_Q = I_B = 14.2~\textrm{A}$, which defines the starting point for the transformation of BEC.

The endpoint for the transformation is set to $I_Q = 0.58~\textrm{A}, I_B = 1.9~\textrm{A}$ which corresponds to a much weaker trap $\{\omega_x,\omega_r\}\approx2\pi\{5.8,15.5\}\textrm{Hz}$, inverted in aspect ratio and displaced (trap minimum) by $\sim9$~mm.
In operation, the trap currents are controlled with a constant current supply with input $V_1$ controlling the current that is passed through both the (\textit{quadrupole}) and (\textit{bias}) pair along with $V_2$ which adds additional current to the (\textit{bias}) pair. 
Thus the coil currents relate to the control inputs by $I_Q(t) \propto V_1(t), I_B(t) \propto V_1(t) + V_2(t)$. 
For each experiment it is these voltage waveforms that are controlled by the ML algorithm. 
Constraints are set to the control parameters based on the minimum quadrupole currents required for trapping atoms and the maximum operating limit of the trap coils.

\subsection*{Outcoupling Procedure}
\label{sec:out_proc}
A key part of our experimental procedure is the ability to measure the amplitude of 3D trap oscillations in a single run of the experiment.  
To achieve this we use the unique detection possibilities afforded by He* and employ a pulsed atom laser \cite{Manning:10,2017arXiv171108886H}, outcoupling multiple pulses over many trap oscillation periods.  
Most standard cold atoms detection schemes would need to vary the switch-off time of the trap over many experimental cycles \cite{PhysRevLett.98.063201,SchaffSongCapuzziEtAl2011}, meaning a cost function could only be generated and fed back to the optimization routine after many experimental runs slowing a comparable experiment considerably. 

We apply a sequence of short (3~$\mu$s) pulses of radio-frequency (RF) radiation to the atoms. 
These pulses are sufficiently Fourier broadened in frequency to uniformly couple the trapped $m_J=+1$ to the magnetically insensitive $m_J=0$ state across the entire BEC.  
Only $\sim 2~\%$ of the trapped atoms are released by each pulse. 
The out-coupled atoms then fall under gravity and are detected on a multi-channel plate and delay line detector \cite{Manning:10} situated 852~mm below the trap center.  
Due to the large ($\sim 20$~eV) internal energy of helium in $2^3S_1$ metastable state, individual atoms are detected with temporal and spatial resolution of $\sim 120$~\textmu{}m in $x$, $y$ and $\sim 10$~\textmu{}m ($\sim 3$~\textmu{}s) in the $z$ (time of flight) directions \cite{2017arXiv171108886H} with a detection quantum efficiency of ${\sim} 10~\%$.  
The results of this procedure are shown in SI Fig.~\ref{fig:fig_si2}.

By binning each out-coupled pulse of atoms (see SI Fig.~\ref{fig:fig_si2}), the mean position and standard deviation may be calculated for each pulse.  
Since only a small fraction of the cloud (100 to 400 atoms) is removed by each pulse, this process can be repeated multiple times (up to 250 pulses in the experiments described in this paper) to determine the COM oscillation along with the any variation in width (breathing modes).  
The time between RF pulses (10~ms) is chosen to be much larger than the temporal width of the pulse and also so that the oscillations in $z$ do not exceed half the pulse period.  
This duration is also much less than the Nyquist limit (31~ms) of the tightest trap axis (15~Hz), ensuring that the oscillation amplitude can be reliably extracted. 
In Fig.~\ref{fig:fig5} multiple experiments were interlaced for each sub-figure varying start time of the atom laser pulses in sub pulse period increments to produce higher temporal resolution.

In this work we encode the cost function as the sum of scaled energies, as a practically realizable proxy for the fidelity of the transformation relative to the adiabatic case.  
In broad terms the cost function is the summation of the atom laser pulse width to the power of four summed with the oscillation width. The pulse width term is heavily penalized by the strong power as the experimental utility of a cold but oscillating BEC is far greater than a thermal cloud with small COM motion. Further (robust) terms are added to prevent erroneously low values of the cost function in `edge' cases (such as when oscillations are larger than the detector) while having no affect in normal operation. For details on the robust cost function see SI.

For each dataset the optimization is first carried out for long duration transformations. It was found that if the algorithm was started 'blind' with no valid cost function such as at short transformation times that it was unable to find the relatively small region of parameter space required to proceed a valid cost function (see Fig \ref{fig:fig4}). At longer times however this region is far larger, so by first optimizing at a longer transformation times and then using the optimized ramp as the initial condition for the next shorter duration one the optimization was able to all but guarantee convergence. It should be noted that like all empirical optimization algorithms ours is unable to guarantee convergence to a `disconnected' optima, in this case particularly those that are not well connected to the slow transformation optima.

\subsection*{Machine Learning Algorithm}

The algorithm used to implement the ML is based on M-LOOP: the Machine-Learning Online Optimization package \cite{Mloop}.  
M-LOOP consists at the basic level of two different processes: a differential evolution (DE) and a Gaussian Process (GP). 
The algorithm starts by executing a set of initial training runs, chosen randomly around the initial experimental parameters provided. 
Once enough data is taken, the GP is then fit to the measured parameters including costs and uncertainties $\{(\mathbf{X}_1,C_1), (\mathbf{X}_2,C_2), \ldots, (\mathbf{X}_N,C_N)\}$ to produce an estimate of the mean cost $C(\mathbf{X})$ and the uncertainty $\sigma_C(\mathbf{X})$, as described in \cite{Wigley2016}. 
The M-LOOP algorithm has been improved since previous work \cite{Wigley2016}, with a key difference being is how the new points ($\mathbf{X}'$) are chosen; points are now selected in a cycle of 5, with the first four chosen by minimizing a biased cost function $B_C(\mathbf{X}) = C(\mathbf{X}) -  b\sigma_C(\mathbf{X})$, where $b$ iterates through $0,1,2$ to $3$. 
When $b$ is 0, the current best estimate of cost parameters is tested, when $b \ne 0$ points with the best potential cost are tested, with increasing risk. 
The data is all normalized, so going beyond $3 \sigma$ was considered an unnecessary risk. 
The fifth step in the cycle is then sampled from an independent differential evolution (DE) algorithm \cite{Storn1997,5601760}. 
This 5th step means some parameters picked for testing are independent of the GP model, ensuring stability in the fitting process.  
This process can be seen in Fig.~\ref{fig:fig3}. 
 All optimizations were carried out with 200 experimental evaluations, as longer evaluations were seen to only yield marginal improvements in cost function while requiring far longer times between experiments to produce the next guess due to complexity scaling of the GP model. In future work alternative methodologies to create similar internal models with superior scaling properties may be investigated \cite{2018arXiv180500654T}.

\phantomsection
\section*{Acknowledgments} 
We thank Marcus Doherty for careful reading of the manuscript.  This work was supported through Australian Research
Council (ARC) Discovery Project Grant No. DP160102337.
S. S. H. is supported by ARC Discovery Early Career Researcher Award No. DE150100315.
D.K.S. is supported by an Australian Government Research Training Program (RTP) Scholarship.

\section*{References}
\phantomsection
\bibliographystyle{apsrev4-1} 
\bibliography{bibliography}


\section*{Supporting Information (SI).}

\subsection{Terminology}
The general technique of empirical optimization has been called many different names   \textit{optimal quantum control} \cite{PhysRevLett.112.240503}, \textit{closed-loop learning} \cite{0953-4075-40-18-R01,doi:10.1063/1.3103486}, \textit{closed-loop learning control} \cite{doi:10.1063/1.3103486,5676677},\textit{ closed-loop optimization} \cite{PhysRevLett.112.240504} and \textit{machine-learning online optimization} \cite{Wigley2016}. While not restricted to quantum mechanical problems these techniques are particularly suited to the high degree of complexity that is often present in such experiments.

\subsection{Adiabatic Timescale for Transport}
The transport present in our experiment also gives rise to an independent adiabatic timescale as shown in the main text for the decompression. Currently in the literature a process is considered non-adiabatic if the duration of the control is shorter than, or comparable to, the trap period $t_{f}^{(ad)}=2\pi/\omega$ \cite{CouvertKawalecReinaudiEtAl2008} which in our system corresponds to 16~ms for the $x$ axis. This is however an incomplete picture as it indicates that the distance traveled has no bearing on the adiabatic timescale. This is made even more puzzling when one considers the acceleration needed to cover some distance in a given time. 

We therefore use the timescale derived in \cite{PhysRevA.89.063414} for transport in a constant trap frequency; which sets the minimal, time average, potential energy, across the transport, to be equal to the first excited state in the final trap.
\begin{equation}
t_{f}^{(ad)}=\sqrt[4]{\frac{4 m d^2}{\hbar \omega_f^3}}
\end{equation}
This then corresponds to 0.27 (0.11)~s for the $x$ axis of our experiment if the final or initial trap frequency is used respectively. The fastest transport shown here thus beats this timescale by a factor of 6.3 (2.5). A full treatment could derive the adiabatic timescale for simultaneous transport and decompression following the work of \cite{BornFock,Jansen}, however this is beyond the scope of this work.

\begin{table}
\centering
\begin{tabular}{lrrrrr}
Experiment & X(mm) & Y(mm) & Z(mm) & E($10^{-32}$~J) & N($10^{3}$)\\
\midrule
Initial & 14.2 & 1.6 & 3.2 & 71.2 & 22.6\\
Damped & 2.8 & 1.7 & 4.0 & 8.9 & 19.7\\
\bottomrule
\end{tabular}
\caption{Detailed damping Results.The system produces strong damping of the original oscillations. There is a small increase in oscillation in the Z axis. We attribute this to the algorithm redistributing oscillations to reduce the cost function.}
\end{table}

\begin{figure}
\includegraphics[width=\linewidth]{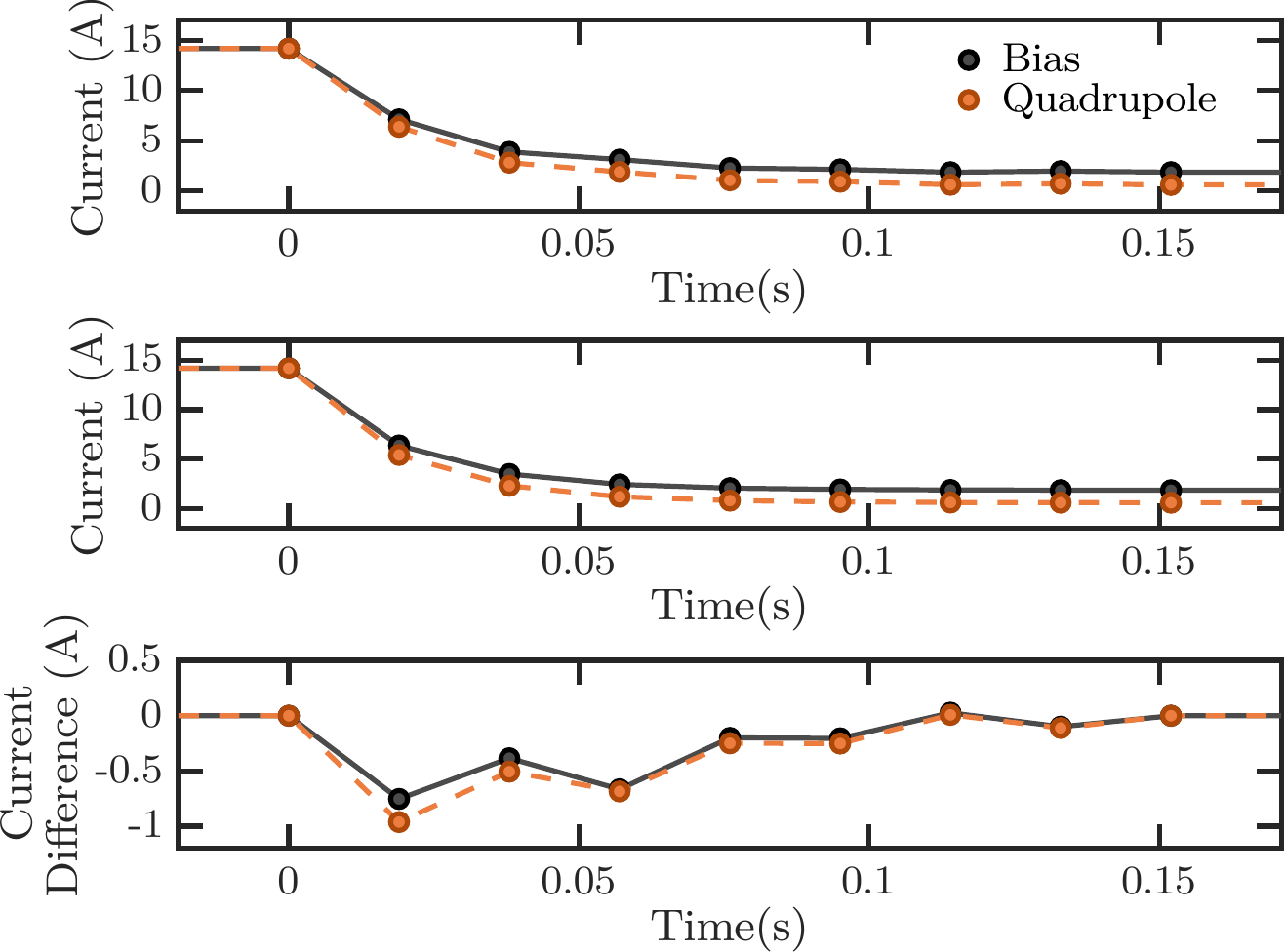}
\caption{ 
The current ramps during the transformation depicted in fig \ref{fig:fig5}. The 14 parameter control (top) is seen to contain non monotonic behavior compared (bottom) to the exponential control (middle). }
\label{fig:fig_si1}
\end{figure}

\begin{figure*}
\centering
\includegraphics[width=\textwidth]{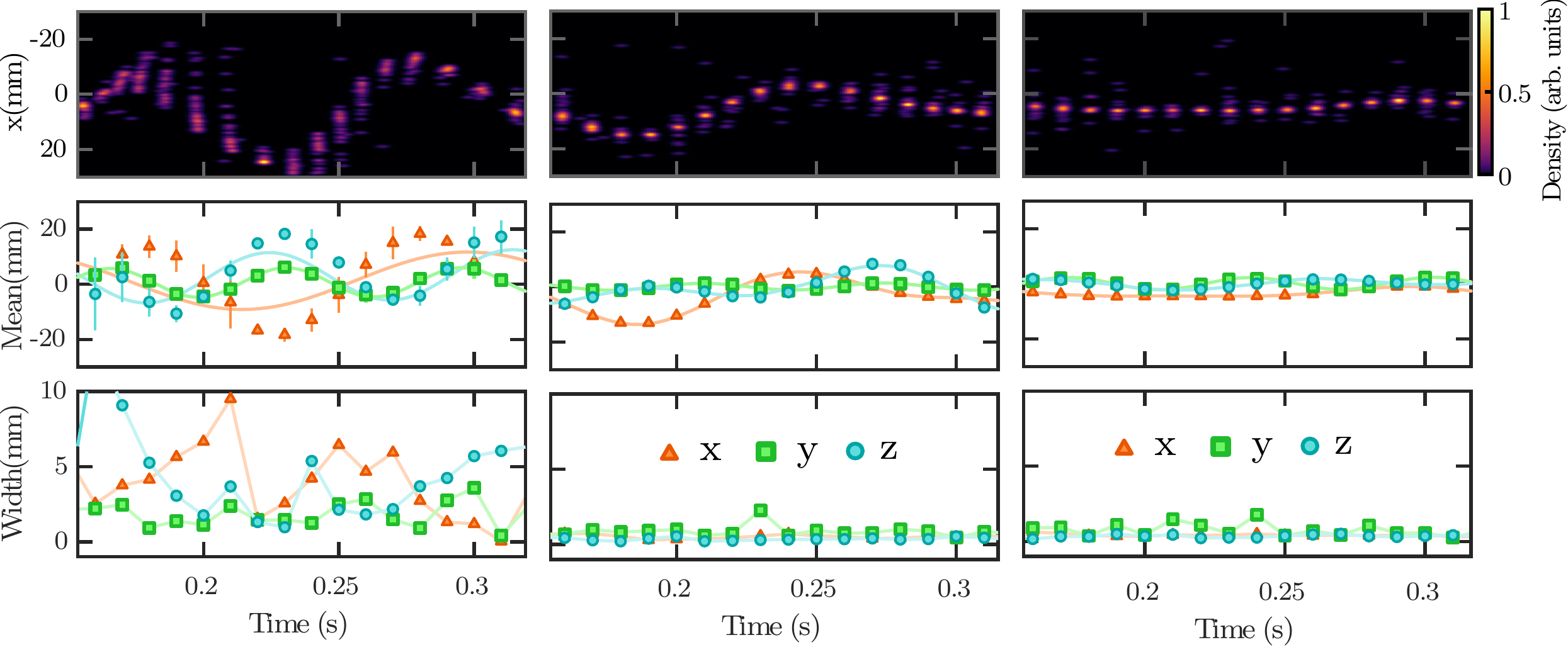}
\caption{Oscillations and heating as measured by the pulsed atom laser in various realizations of the 153.6ms control duration experiment. 
Left:A transformation that results in large heating and oscillation dephasing through anharmonics of the trap potential. 
Center:large oscillations without heating. 
Right: the oscillations after the transformation depicted in \ref{fig:fig5}(a) and the optimum for this duration. 
Corresponding costs are 0.579,0.0412 and 0.0024 (arb. units) for the left,middle and right respectively. 
Top: Histogram of the detected atoms from the pulsed atom laser. Middle: Mean position of each pulse in each Cartesian axis with sinusoidal fits shown with corresponding amplitude in (x,y,z)=(10.2,5.3,8.9),(10.2,1.2,7.0),(2.5,2.2,3.7) for the left,middle and right respectively. Fits are performed to the full atom laser pulse sequence which contains 250 pulses, 15 of which are shown here.
Note the phase modulation due to oscillations in the z(time) axis. }
\label{fig:fig_si2}
\end{figure*}

\begin{figure}
\includegraphics[width=250px]{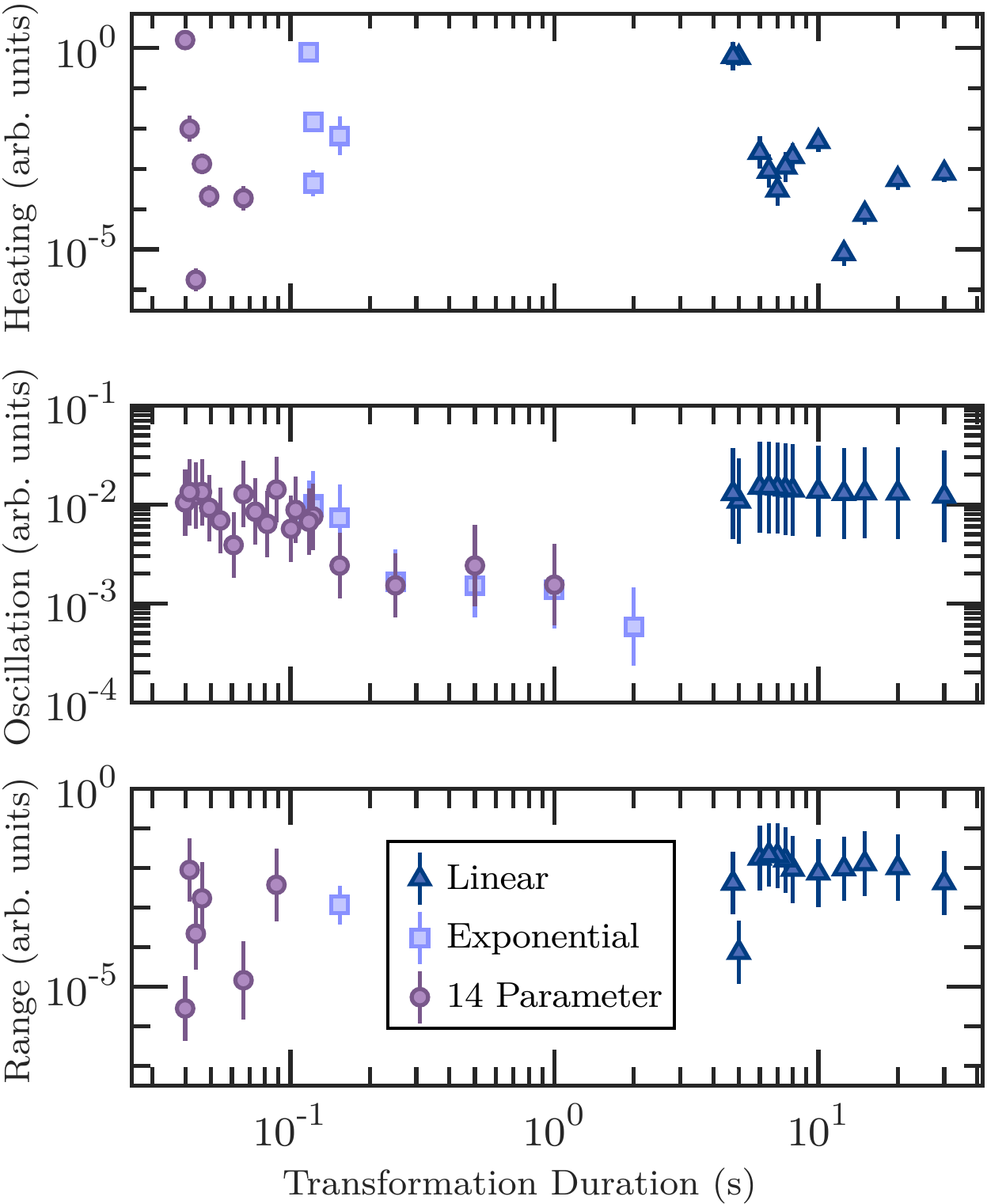}
\caption{Component costs that make up the minimum cost functions in a optimization sequence Fig.~\ref{fig:fig2}. The dramatic increase in the heating penalty at low transformation durations is apparent. No data points have any associated number penalty. Legend is common across sub figures.}
\label{fig:fig_si3}
\end{figure}

\subsection{Robust Cost} \label{SI:robust_cost}
The development of a robust cost function that encodes the users desire while only accessing  limited, noisy information is a general problem in optimization. To a first approximation we wish to maximize the fidelity of the final state with respect to the state produced though an adiabatic transformation. However measuring this quantity experimentally is impractical and importantly even if possible, it is unclear how one would produce the `reference' adiabatically transformed state as this is the goal of the optimization. Further in practical usage a small COM motion of the final state is far more tolerable for use in experiments \cite{Khakimov2016} than the equivalent decrease in fidelity caused by heating or atom loss. 

To overcome both the cost function used here is based on a measurement of the total energy of the transformed state. As the desired state has the minimum total energy we can use the minimization thereof as a proxy to produce the maximum fidelity. Additionally we may manipulate the weighting of the COM energy against temperature to reflect the relative experiment tolerance. 

 To calculate the cost we utilize the pulsed atom laser method described in the main text. We bin the continuous detected count data into atom laser pulse $i$ comprising detections $j$ with spatial coordinates $\mathbf{D}_{ij}=(x_{ij},y_{ij},z_{ij})$. To convert a detections time $t$ to the spatial value $z_{ij}$ we subtract the time of the corresponding out-coupling pulse along with free fall time to the detector, this is then multiplied by the velocity at the detector. The cost is then comprised primarily of two elements:
 \begin{equation}
C_{core}=\lvert\sigma_{i}(\mu_{j}(\mathbf{D}_{ij}))\rvert+\mu_{i}(S(\lvert \sigma_j(\mathbf{D}_{ij})\rvert))
\label{eq:cost_basic}
\end{equation}
where  $\mu_{k}$ ($\sigma_{k}$) is the mean (standard deviation) over index k, S is the width scaling function and $\lvert\ \rvert$ is the vector norm. The first term (henceforth oscillation cost) is proportional to the the standard deviation of the center of mass of the BEC. The second, the mean scaled width of each pulse, is an indirect measure of the BEC temperature (henceforth width cost).  Notably without this term algorithm is prone to producing a thermal cloud that quickly damps out its COM motion in order to minimize the first term.

The scaling function has three motivations: to prevent any penalization of the mean field energy and in turn atom number in the final state, to increase the penalty for a thermal gas compared to a BEC, and finally to prevent any discontinuities in the cost function that could be complex to represent in the internal GP model. 
\begin{equation}
	S(x)=\\
   \begin{cases}
     0, &A(i)<H_w \\
     P_w\times(x - H_w)^4, &A(i)\geq H_w
   \end{cases}
\label{eq:width_scaling}
\end{equation}
Here $H_w$ denotes the threshold for this width scaling and $P_w$ the proportionality constant. For this experiment $H_w=4$~mm was chosen to be slightly less than the widths observed from the linear ramps the proportionality $P=6*10^{8}m^{-4}$ set such that the hottest clouds (eg. Fig. \ref{fig:fig_si2} left) have cost $\sim0.5$. 

 While this cost function captures the majority of the relevant physics optimization algorithms can quickly find regions where the assumptions used in its construction do not hold. In this work namely when oscillations are so large that the atom pulses only occasionally 'sweep' across the detector.
 
For the oscillation cost is modified by adding a piecewise power function to $\sigma_{i}(\mu_{j}(\mathbf{D}_{ij}))$ which above threshold strongly penalizes large peak to peak amplitudes while is zero below. The peak to peak oscillation is not used as the main cost function due to its higher noise.
The the cost is thus modified as
 \begin{equation}
C_{core}=\lvert\sigma_{i}(\mu_{j}(\mathbf{D}_{ij}))+R(\mu_{j}(\mathbf{D}_{ij}))\rvert+\mu_{i}(S(\lvert \sigma_j(\mathbf{D}_{ij})\rvert))
\label{eq:cost_robust}
\end{equation}
where R(x) is a weighting against the range (peak to peak) of the oscillation. The form that is used here is
\begin{equation}
	R(x)=
   \begin{cases}
     0, &Range_{i}(x)<H_o \\
     P_{o}\times(Range_{i}(x)- H_o)^4, &Range_{i}(x)\geq H_o
   \end{cases}
\label{oscAC}
\end{equation}
where $H_o$ denotes the threshold for this oscillation scaling and $P_o$ the proportionality constant. For this experiment $H_o=47$~mm was chosen as size of the low noise region of our detector and the proportionality $P_{o}=1*10^{6}m^{-4}$ set such that oscillations that reach the noise region have cost $\sim0.03$. 

Not all transformations resulted in a valid cost function to report back to the ML algorithm namely, there were many cases when the minimum number of atoms were not detected. In this case we returned the failed evaluation condition back to the ML algorithm. For this work the threshold for this condition was set to 2500 detected atoms.

\end{document}